# Determinação do parâmetro de relaxação ótimo num procedimento numérico de propagação de sólitons

**Eliandro Rodrigues Cirilo, Paulo Laerte Natti, Neyva Maria Lopes Romeiro**

Departamento de Matemática, Universidade Estadual de Londrina

86051-990 Londrina, PR

**Érica Regina Takano Natti**

Pontifícia Universidade Católica do Paraná, Campus Londrina

86060-000 Londrina, PR

Resumo: Neste trabalho, considerando um procedimento numérico desenvolvido para resolver um sistema de equações diferenciais acopladas, complexas e não-lineares, que descreve a propagação de sólitons em fibras óticas dielétricas, otimizamos o tempo de processamento numérico, em relação ao parâmetro de relaxação do procedimento, para conjuntos relevantes de valores das variáveis dielétricas da fibra ótica.

Palavras-chave: sóliton ótico, tempo de processamento, otimização.

Abstract: In this work, considering a numerical procedure developed to solve a system of coupled nonlinear complex differential equations, which describes the solitons propagation in dielectric optical fibers, we optimize the numerical processing time, in relation to the relaxation parameter of the procedure, for relevant groups of values of the dielectric variables of the optic fiber.

## 1 Introdução

O século XX pode ser chamado de era da física linear, pois foi dominado por equações lineares (Maxwell, Schrödinger...), por objetos matemáticos lineares (espaços vetoriais, em particular os espaços de Hilbert), e por métodos lineares (transformada de Fourier, teoria de perturbação...). Naturalmente a importância da não-linearidade, começando com as equações de Navier-Stokes e passando para as teorias da gravitação geral e dos campos quantizados com aplicações em física dos sólidos, física nuclear e em física de partículas, foi reconhecida, embora o tratamento destes efeitos fosse muito difícil, exceto como perturbações da solução básica da teoria linearizada (NATTI; PIZA, 1996).

Durante as quatro últimas décadas, tornou-se cada vez mais evidente que não-linearidades podem resultar em novos fenômenos, os quais não podem ser obtidos via teoria de perturbação. Este é o caso de ondas do tipo sóliton (EILENBERGER, 1981; TAYLOR, 1992).

Sólitons representam fenômenos que apresentam as características de serem não-lineares, localizados, quase-estáticos e que interagem fortemente mantendo sua identidade. Uma grande variedade de fenômenos apresenta tais propriedades. Citamos algumas áreas em física e áreas correlatas, nas quais a teoria de sólitons tem sido utilizada com freqüência: física de plasma, física do estado sólido, física das partículas elementares, meteorologia, e em particular, para descrever fenômenos não-lineares em ótica.

Devido à sua estabilidade, os sólitons podem ser transmitidos em fibras óticas por longas distâncias. Hasegawa e Tappert (1973) foram os primeiros a sugerirem que sólitons poderiam se propagar em fibras óticas, devido ao equilíbrio entre a auto-modulação de fase e a dispersão no regime anômalo, contudo foi somente em 1980 que tais efeitos foram verificados experimentalmente (MOLLENAUER; STOLEN; GORDON, 1980). Posteriormente, Hasegawa (1984) propôs que os sólitons pudessem ser transmitidos para aumentar o desempenho das telecomunicações óticas. Emplit et al. (1987) foram os primeiros a realizarem a observação experimental da transmissão em picosegundos de sólitons em fibras óticas. No ano seguinte, Mollenauer et al. (1988) transmitiram pulsos de sólitons por 4000 quilômetros usando o efeito Raman. Uma equipe da Bell Labs em 1991 propagou sólitons por mais de 14.000 quilômetros utilizando uma fibra ótica dopada com érbio (MOLLENAUER et al., 1991). Em 1998, uma equipe da France Telecom coordenada por Thierry Georges, combinando sólitons óticos de diferentes comprimentos de onda, transmitiu dados à taxa de 1 terrabit por segundo (LE GUEN et al., 1999). Enfim, o uso prático de sólitons transformou-se em realidade quando a Algety Telecom desenvolveu, em 2001, equipamentos de telecomunicações submarinos, na Europa, transmitindo sólitons.

O estudo teórico das propriedades dielétricas que uma fibra ótica deve apresentar, de modo que a comunicação via sólitons seja mais estável e eficaz do que a comunicação via sinais lineares no formato NRZ (padrão atual), é um tema de grande importância tecnológica e econômica, com aplicações potenciais em Ciências dos Materiais e das Telecomunicações. Na última década tem sido abordado com intensidade estudos, teóricos e experimentais, sobre temas relacionados com os processos de geração (MALOMED et al., 2005), propagação (HASEGAWA, 2000; GALLEAS et al., 2003) e estabilidade (CHEN; ATAI, 1998; YMAI et al., 2004) de sólitons em fibras dielétricas.

Neste contexto de comunicação ótica, em (GALLEAS et al., 2003; YMAI et al., 2004) estudamos analiticamente a propagação e a estabilidade de sólitons em guias dielétricos ideais, enquanto que em (QUEIROZ et. al. 2006; CIRILO; ROMEIRO; NATTI, 2007) desenvolvemos procedimentos numéricos para realizar tais estudos. Nos próximos trabalhos, pretendemos descrever o comportamento de sólitons em guias não-ideais, ou ainda, como a propagação e a estabilidade de ondas tipo sólitons são afetadas quando introduzimos perturbações, ou termos perturbativos, nas EDP's estudadas em YMAI et al. (2004). Citamos como possíveis processos perturbativos locais que afetam a propagação de sóliton em guias dielétricos não-ideais: (i) absorções de várias espécies tais como inomogeneidades, difusão de moléculas de hidrogênio e bolhas na fibra (RAGHAVAN; AGRAWAL, 2000); (ii) defeitos na fabricação da fibra ótica tais como variações no diâmetro da fibra, rugosidade, sinuosidade no eixo longitudinal, micro curvaturas, emendas por máquina de fusão de arco voltaico (STROBEL, 2004); (iii) irradiação devido à curvatura da fibra; (iv) ruídos nos campos elétricos das ondas do pacote gerado, por exemplo, pelo processo de bombeamento do sóliton com o objetivo de compensar a absorção da fibra ótica (WERNER; DRUMMOND, 1993); entre outros processos perturbativos. Note que neste nível soluções analíticas não são conhecidas, de modo que os estudos deverão ser realizados através de procedimentos numéricos.

Com o objetivo de estudar a estabilidade de sólitons em guias não-ideais, já desenvolvemos e validamos em (CIRILO; ROMEIRO; NATTI, 2007) um procedimento numérico geral, aplicando-o ao sistema de EDP's que descreve a propagação de sólitons em guias ideais, situação em que conhecemos a solução analítica (GALLEAS et al., 2003). A partir deste procedimento, será possível estudar a solução numérica de sistemas de EDP's mais complexos, que descrevem de forma realista a propagação de sólitons em materiais dielétricos.

Neste trabalho propomos otimizar no procedimento numérico descrito em (CIRILO; ROMEIRO; NATTI, 2007) o tempo de processamento numérico, com relação ao parâmetro de relaxação, para conjuntos relevantes de valores das variáveis dielétricas da fibra ótica. Em De Oliveira et al. (2007) verificamos que o esquema numérico desenvolvido, mostrou-se mais sensível às variações de um parâmetro, que é uma medida da intensidade da não-linearidade do sistema de EDP's. Quando aumentamos a intensidade dos efeitos não-lineares da fibra ótica, o procedimento desenvolvido propaga erros com maior intensidade, fazendo com que a solução numérica não convirja para a solução analítica, para uma dada discretização da malha. Por outro lado, verificamos que ao refinar a malha, podemos controlar estes erros numéricos, contudo o tempo de processamento numérico cresce. Sabendo que ao introduzirmos efeitos perturbativos em nosso modelo estaremos aumentando o número de operações numéricas, torna-se essencial otimizar o tempo de processamento em nosso procedimento. Na seção 2 apresentamos a solução analítica sóliton de um sistema de EDP´s que descreve a propagação de pacotes de ondas em fibras óticas ideais do tipo $\chi^{(2)}$. Na seção 3 descrevemos o nosso procedimento numérico e justificamos a necessidade da otimização do tempo de processamento numérico, em relação ao parâmetro de relaxação do procedimento. Na seção 4 apresentamos os resultados do processo de otimização numérica. Enfim, na seção 5, resumimos os resultados obtidos.

## 2 Dinâmica longitudinal de campos acoplados em dielétricos tipo $\chi^{(2)}$

Nesta seção apresentamos o sistema de EDP's acoplado, não-linear e complexo, obtido a partir das equações de Maxwell, que descreve a propagação longitudinal de duas ondas eletromagnéticas acopladas (modos fundamental e segundo harmônico) em fibras óticas dielétricas ideais, de seção transversal retangular, com não-linearidades do tipo $\chi^{(2)}$ (AGRAWAL, 1995). A modelagem matemática deste sistema de EDP's é realizada de modo detalhado em Galleas et al. (2003) e dadas pelas equações

$$I\frac{\partial a_1}{\partial \xi} - \frac{r}{2}\frac{\partial^2 a_1}{\partial s^2} + a_1^* a_2 \exp(-I\beta\xi) = 0$$
$$I\frac{\partial a_2}{\partial \xi} - I\delta\frac{\partial a_2}{\partial s} - \frac{\alpha}{2}\frac{\partial^2 a_2}{\partial s^2} + a_1^2 \exp(I\beta\xi) = 0 \qquad (1)$$

onde $I = \sqrt{-1}$ é a unidade imaginária e $a_1(\xi,s)$ e $a_2(\xi,s)$ são variáveis complexas que representam as amplitudes normalizadas dos campos elétricos das ondas fundamental e segundo harmônico, respectivamente. Em (1) a variável independente s tem caráter de dimensão espacial, a direção unidimensional da propagação da onda sóliton, enquanto a variável independente $\xi$ tem caráter temporal (GALLEAS et al., 2003).

O sistema (1) apresenta soluções analíticas do tipo sólitons para certos domínios de valores dos parâmetros $\alpha$, $\beta$, $\delta$ e r (YMAI et al., 2004). Esses parâmetros caracterizam as propriedades dielétricas da fibra ótica. No limite $|\beta| \to \infty$, as equações diferenciais (1) se desacoplam na equação de Schrödinger não-linear (MENYUK; SCHIEK; TORNER, 1994), correspondendo ao limite de incompatibilidade de fase (ondas fora de fase). Verifica-se que a quantidade $\beta$ mede a taxa de geração do segundo harmônico, ou ainda, que é uma medida da intensidade de não-linearidade do material, já que o termo não-linear em (1) é responsável pelo acoplamento dos campos elétricos das ondas fundamental e segundo harmônico. A quantidade r é um indicador do regime de dispersão da onda fundamental. Quando temos $r = +1$, a onda fundamental encontra-se no regime de dispersão dito normal, porém se $r = -1$, a onda fundamental encontra-se no regime de dispersão dito anômalo. A quantidade α mede a dispersão relativa das velocidades de grupo das ondas fundamental e segundo harmônico no material (fibra ótica). Para valores de $|\alpha| > 1$, a onda do segundo harmônico possui dispersão maior que a onda fundamental e para valores de $|\alpha| < 1$ é a onda fundamental que tem dispersão maior. Analogamente à quantidade r, o sinal da quantidade dielétrica α é um indicador do regime de dispersão, mas da onda do segundo harmônico. Dizemos que para valores de α positivos, a onda do segundo harmônico se encontra no regime de dispersão normal, enquanto que para valores de α negativos, a onda do segundo harmônico se encontra no regime de dispersão anômalo. Enfim, a quantidade δ mede a diferença das velocidades de grupo dos modos fundamental e segundo harmônico, e assume valores não-nulos, por exemplo, em meios anisotrópicos, quando os vetores de Poynting destes modos encontram-se desalinhados (*walk-off wave*). Em Galleas et al. (2003) há uma descrição pormenorizada da interpretação das quantidades dielétricas α, β, δ e r, relacionando-as com as propriedades óticas da fibra. Observamos que podemos escolher as características da onda sóliton a ser propagada na fibra ótica (velocidade, largura, amplitude, estabilidade, etc..), selecionando ou propondo materiais com as propriedades dielétricas α, β, δ e r adequadas.

As soluções analíticas tipo sólitons das equações (1) são conhecidas (GALLEAS et al., 2003) e dadas por

$$a_1 = \pm \frac{3}{2(\alpha - 2r)} \sqrt{\alpha r \left( \frac{\delta^2}{2\alpha - r} + \beta \right)} \operatorname{sch}^2 \left[ \pm \sqrt{\frac{1}{2(2r - \alpha)} \left( \frac{\delta^2}{2\alpha - r} + \beta \right)} \left( s - \frac{r\delta}{2\alpha - r} \xi \right) \right] \times$$

$$\exp \left\{ I \left[ \frac{r\delta^2 (4r - 5\alpha)}{2(2\alpha - r)^2 (2r - \alpha)} - \frac{r\beta}{2r - \alpha} \right] \xi - \frac{I\delta}{2\alpha - r} s \right\} \quad (2)$$

$$a_2 = \frac{3r}{2(\alpha - 2r)} \left[ \frac{\delta^2}{2\alpha - r} + \beta \right] \operatorname{sch}^2 \left[ \pm \sqrt{\frac{1}{2(2r - \alpha)} \left( \frac{\delta^2}{2\alpha - r} + \beta \right)} \left( s - \frac{r\delta}{2\alpha - r} \xi \right) \right] \times$$

$$\exp \left\{ 2I \left[ \frac{r\delta^2 (4r - 5\alpha)}{2(2\alpha - r)^2 (2r - \alpha)} - \frac{r\beta}{2r - \alpha} + \frac{\beta}{2} \right] \xi - \frac{2I\delta}{2\alpha - r} s \right\} \quad (3)$$

Em Queiroz et. al. (2006) adaptamos um procedimento numérico, desenvolvido anteriormente por um dos autores, que forneceu a solução numérica de (1) no caso em que $\delta = 0$. No caso da propagação de sólitons em fibras óticas ordinárias, e em situações ótimas, o fenômeno *walk-off wave* pode ser desconsiderado (ARTIGAS; TORNER; AKHMEDIEV, 1999), o que justifica tomar $\delta = 0$ nestas situações. Contudo, ao estendermos este procedimento numérico a fibras óticas não-ideais, teremos, necessariamente, $\delta \neq 0$, o que justifica o desenvolvimento de um procedimento numérico geral adaptado ao sistema (1).

## 3 Modelo numérico

O esquema numérico adotado neste trabalho foi aproximar as derivadas por diferenças finitas e resolver o sistema algébrico resultante da discretização, implicitamente, através do método de Gauss-Seidel com relaxação (SMITH, 1990; SPERANDIO, 2003).

O modelo de ondas tipo sóliton é resolvido no domínio dado por $\xi \times s = [0, T] \times [-L, L]$, onde $T, L \in \Re$. Identificamos as variáveis $a_1(\xi, s) \equiv a_1(k+1, i)$ e $a_2(\xi, s) \equiv a_2(k+1, i)$, onde consideramos que $k = 0, 1, ..., k_{max}$ e $i = 1, 2, ..., ni$, onde $k_{max}$ é o último avanço em $\xi$ e $ni$ o número máximo de pontos em $s$. O domínio fica definido por uma malha computacional discreta de $k_{max} \times ni$ pontos, como apresentado na figura 1.

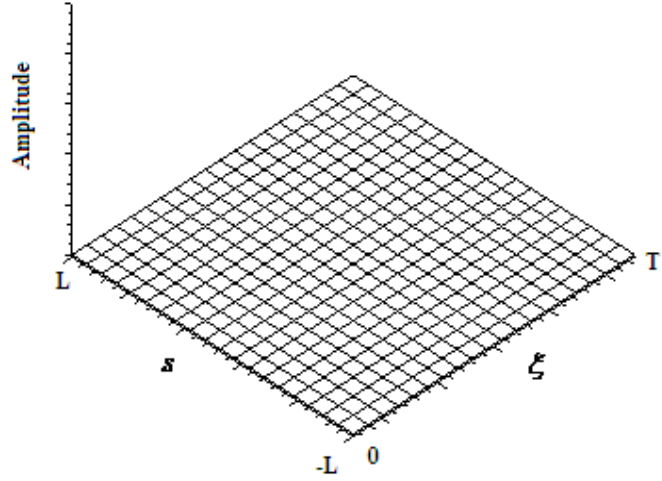

Figura 1: Domínio computacional para a propagação da onda sóliton

Aproximando as derivadas temporais $\frac{\partial a_1}{\partial \xi}$ e $\frac{\partial a_2}{\partial \xi}$ por diferença progressiva, e por diferença central as derivadas espaciais $\frac{\partial a_2}{\partial s}$, $\frac{\partial^2 a_1}{\partial s^2}$ e $\frac{\partial^2 a_2}{\partial s^2}$, obtivemos a partir das equações (1) os seguintes sistemas lineares

$$a_1(k+1,i)^{n+1} = \left(\frac{1}{{}^1A_p}\right)\left[ {}^1A_W a_1(k+1,i-1)^{n+1} + {}^1A_E a_1(k+1,i+1)^n + {}^1A_{p_0} a_1(k,i) \right.$$
$$\left. - a_1^*(k,i) a_2(k,i) \exp(-I\beta\xi) \right]$$

$$a_2(k+1,i)^{n+1} = \left(\frac{1}{{}^2A_p}\right)\left[ {}^2A_W a_2(k+1,i-1)^{n+1} + {}^2A_E a_2(k+1,i+1)^n + {}^2A_{p_0} a_2(k,i) \right.$$
$$\left. - \left(a_1(k+1,i)^{n+1}\right)^2 \exp(I\beta\xi) \right]$$

onde

$${}^1A_p = \frac{I}{\Delta\xi} + \frac{r}{(\Delta s)^2} \qquad {}^1A_E = {}^1A_W \qquad {}^1A_W = \frac{r}{2(\Delta s)^2} \qquad {}^1A_{p_0} = \frac{I}{\Delta\xi}$$

$${}^2A_p = \frac{I}{\Delta\xi} + \frac{\alpha}{(\Delta s)^2} \qquad {}^2A_E = \frac{I\delta}{2\Delta s} + \frac{\alpha}{2(\Delta s)^2} \qquad {}^2A_W = -\frac{I\delta}{2\Delta s} + \frac{\alpha}{2(\Delta s)^2} \qquad {}^2A_{p_0} = \frac{I}{\Delta\xi}$$

que reescritos em termos do parâmetro de relaxação w tomam a forma

$$a_1(k+1,i)^{n+1} = w\, a_1(k+1,i)^{n+1} + (1.0-w)\, a_1(k+1,i)^n \qquad (4)$$

$$a_2(k+1,i)^{n+1} = w\, a_2(k+1,i)^{n+1} + (1.0-w)\, a_2(k+1,i)^n \qquad (5)$$

sendo $0 < w < 2$ e n o nível iterativo de cálculo (SMITH, 1990).

Impomos como condições iniciais de (4-5) as soluções analíticas dadas em (2-3), ou seja, $a_1(0,i)$ e $a_2(0,i)$. Para as condições nos contornos, admitindo-se que L seja suficientemente grande, impomos que

$$a_1(k+1,\ 1) = 0,\ a_1(k+1, ni) = 0 \qquad e \qquad a_2(k+1,\ 1) = 0,\ a_2(k+1, ni) = 0 \qquad (6)$$

Os sistemas lineares complexos (4-5), em conjunto com as condições iniciais e de contornos dadas, foram resolvidos iterativamente através do método de Gauss-Seidel com relaxação, impondo um fator de tolerância para o erro de $1\times 10^{-6}$. O código foi desenvolvido em Fortran. Utilizamos uma CPU com processador Intel Celeron M 420 com 1.6 GHz, 533 MHz FSB e 512 MB DDR2. Nos cálculos desta seção, sempre, tomaremos $L = 30$ e $T = 10$, fixando assim as dimensões da malha.

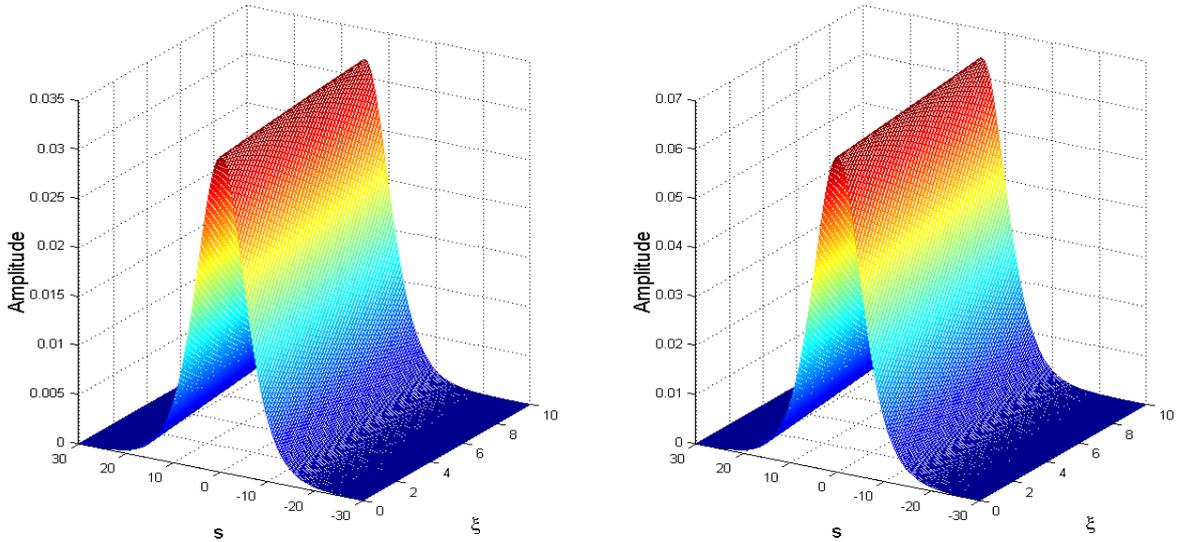

Figura 2: Soluções numéricas de $|a_1(s,\xi)|$, à esquerda, e $|a_2(s,\xi)|$, à direita, para $\alpha = -1/4$, $\beta = -0.2$, $\delta = -1/4$ e $r = -1.0$.

Na figura 2 apresentamos a propagação da onda sóliton, calculada numericamente em uma malha com $\Delta s = 2.4\times 10^{-1}$ (250 pontos) e $\Delta\xi = 1.0\times 10^{-2}$ (1000 pontos), quando tomamos para as constantes dielétricas os seguintes valores: $\alpha = -1/4$, $\beta = -0.2$, $\delta = -1/4$ e $r = -1.0$. Na figura 3 são comparados os perfis finais do módulo do harmônico fundamental e segundo harmônico, obtidos numericamente na figura 2, com as soluções analíticas dadas em (2-3). Note que ocorre uma concordância significativa entre os resultados numéricos e analíticos.

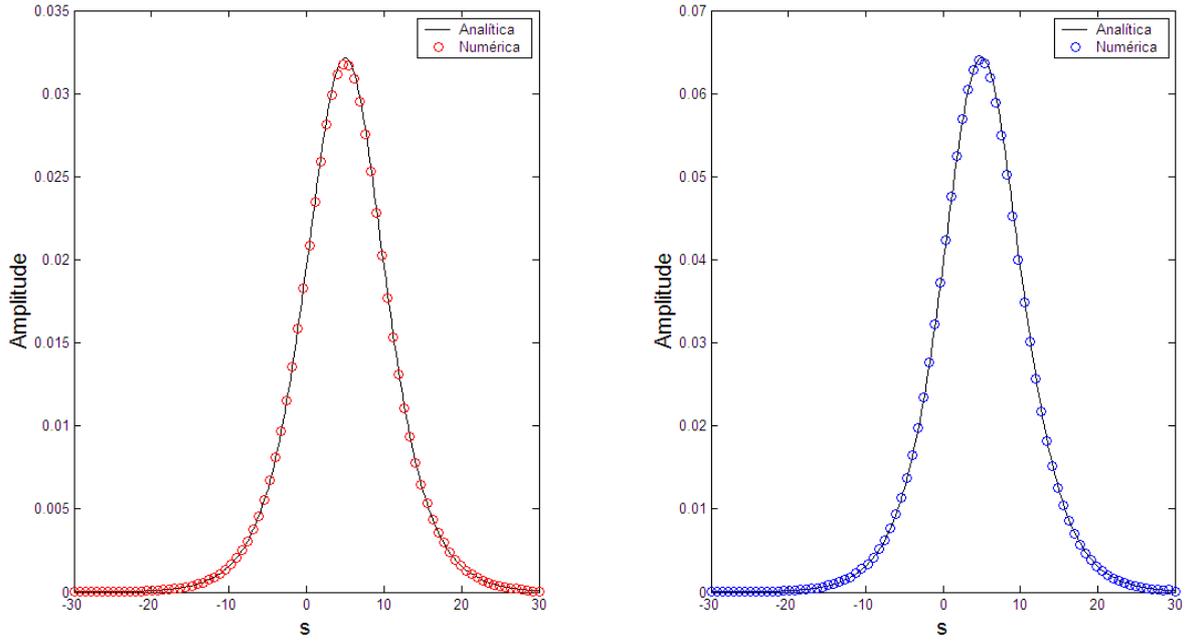

Figura 3: Soluções analíticas e numéricas de $|a_1(s,\xi)|$, à esquerda, e $|a_2(s,\xi)|$, à direita, em $\xi = 10$, para $\alpha = -1/4$, $\beta = -0.2$, $\delta = -1/4$ e $r = -1.0$.

Por outro lado, em De Oliveira et al. (2007) verificamos que nosso desenvolvimento numérico mostrou-se bastante sensível às variações do parâmetro $\beta$, que é uma medida da intensidade da não-linearidade do sistema de EDP's, veja equações (1). Notamos que conforme o valor absoluto de $\beta$ cresce, para que a diferença entre as soluções analítica e numérica seja aceitável, devemos refinar a malha em $\Delta s$ e $\Delta \xi$, o que implicará inevitavelmente num maior tempo de processamento numérico. Além disso, como pretendemos utilizar este desenvolvimento numérico para descrever a propagação de sólitons em fibras óticas reais, onde novos efeitos perturbativos devem ser acrescentados em nosso modelo, o que acarretará no aumento do número de operações numéricas, torna-se essencial otimizar o tempo de processamento numérico em nosso procedimento.

## 4 Otimização numérica do tempo de processamento

A variação do parâmetro de relaxação $w$, em (4) e (5), faz com que o processo de convergência do procedimento numérico seja acelerado ou desacelerado. Logo, existe um valor ótimo para este parâmetro, tal que a velocidade de convergência é máxima, minimizando o tempo de processamento numérico. O valor ótimo do parâmetro é calculado por $w = \dfrac{2}{1+\sqrt{1-\rho^2(B)}}$, onde $\rho(B)$ é o raio espectral da matriz de iteração $B$ de Jacobi. A matriz $B$ esta associada à matriz dos coeficientes do sistema linear a ser resolvido (SMITH, 1990). Em geral é extremamente difícil calcular esse valor para sistemas lineares de pequeno porte. Em sistemas de médio e grande porte, onde o nosso problema reside, pode ser impossível o cálculo analítico deste valor ótimo. Desta forma, por intermédio da experimentação numérica, vamos otimizar o tempo de processamento, em relação ao parâmetro de relaxação do procedimento numérico, para conjuntos de valores da variável dielétrica $\beta$. O parâmetro de relaxação $0 < w < 2$ é uma medida do peso, na solução numérica, entre o método de Gauss-Seidel e a solução do nível iterativo anterior. Em (4) e (5) podemos ter as seguintes situações: se $w = 1.0$ estamos utilizando para o procedimento numérico o método de Gauss-Seidel, por outro lado se $0 < w < 1.0$ dizemos que o procedimento é sub-relaxado - SUR (*Sucessive Under Relaxation*), enquanto que para $1.0 < w < 2.0$ tal procedimento é sobre-relaxado - SOR (*Sucessive Over Relaxation*). Para valores do parâmetro fora da faixa $0 < w < 2$ não há convergência no cálculo da solução numérica (SMITH, 1990).

Fixamos nas simulações desta seção os seguintes valores para as variáveis dielétricas: $\alpha = -1/4$, $\delta = -1/4$, $r = -1.0$ e variamos $\beta$, para $\beta < -1/8$, satisfazendo a condição de existência da solução sóliton (YMAI et al., 2004). Consideramos também uma malha computacional com a variável $s$ no intervalo $-30 \leq s \leq 30$ e uma discretização entre $\Delta s = 2.4 \times 10^{-1}$ (250 pontos) e $\Delta s \approx 0.1 \times 10^{-1}$ (550 pontos), enquanto a variável $\xi$ foi fixada no intervalo $0 \leq \xi \leq 10$ com uma discretização entre $\Delta \xi = 1 \times 10^{-2}$ (1000 pontos) e $\Delta \xi = 1 \times 10^{-3}$ (10000 pontos).

Na tabela 1, variando o parâmetro dielétrico $\beta$ com os demais parâmetros fixados, calculamos o tempo de processamento numérico para propagar a onda sóliton. Para cada conjunto de parâmetros dielétricos fixamos um refinamento da malha, dado na parte inferior da tabela 1, que é definido exigindo-se que o maior erro entre as soluções analíticas e numéricas para $|a_1(s,\xi)|$ e $|a_2(s,\xi)|$ sejam inferiores a $10^{-2}$. Os valores das células em tons de cinza da tabela 1 referem-se aos tempos de computação simulados, em minutos, na CPU descrita anteriormente. As células em tom cinza-claro não apresentam valores para o tempo de processamento, pois os erros máximos nos cálculos de $|a_1(\xi,s)|$ ou $|a_2(\xi,s)|$ foram superiores a $10^{-2}$ para o refinamento escolhido na malha computacional, ou porque o procedimento numérico tendeu a tempos de máquina proibitivos. Os valores das células em tom cinza-médio foram aqueles em que os erros máximos nos cálculos de $|a_1(\xi,s)|$ e $|a_2(\xi,s)|$ não foram superiores a $10^{-2}$ para os refinamentos escolhidos para a malha computacional. Finalmente, os tempos ótimos de processamento são indicados nas células em tom cinza-escuro.

Tabela 1 - Tempo de processamento numérico $T_p$ (minutos) em função dos parâmetros $\beta$ e $w$.

| $T_p$ (minutos) | | $\beta$ | | | | | | |
|---|---|---|---|---|---|---|---|---|
| | | $-0.2$ | $-0.7$ | $-1.2$ | $-1.7$ | $-2.2$ | $-2.7$ | $-3.0$ |
| $w$ | 0.1 | $1.9 \times 10^{-1}$ | $3.7 \times 10^{-1}$ | $4.5 \times 10^{-1}$ | $6.1 \times 10^{-1}$ | | | |
| | 0.3 | $1.1 \times 10^{-1}$ | $1.8 \times 10^{-1}$ | $2.1 \times 10^{-1}$ | $2.8 \times 10^{-1}$ | $6.1 \times 10^{-1}$ | | |
| | 0.5 | $9.1 \times 10^{-2}$ | $1.3 \times 10^{-1}$ | $1.5 \times 10^{-1}$ | $1.9 \times 10^{-1}$ | $4.3 \times 10^{-1}$ | | 2.6 |
| | 0.7 | $8.1 \times 10^{-2}$ | $1.1 \times 10^{-1}$ | $1.2 \times 10^{-1}$ | $1.5 \times 10^{-1}$ | $3.3 \times 10^{-1}$ | 1.9 | 2.2 |
| | 0.9 | $7.1 \times 10^{-2}$ | $9.5 \times 10^{-2}$ | $9.9 \times 10^{-2}$ | $1.22 \times 10^{-1}$ | $2.8 \times 10^{-1}$ | 1.5 | 1.7 |
| | 1.0 | $6.7 \times 10^{-2}$ | $8.3 \times 10^{-2}$ | $9.2 \times 10^{-2}$ | $1.2 \times 10^{-1}$ | $2.5 \times 10^{-1}$ | 1.4 | 1.5 |
| | 1.2 | $7.7 \times 10^{-2}$ | $1.0 \times 10^{-1}$ | $1.1 \times 10^{-1}$ | $1.4 \times 10^{-1}$ | $3.1 \times 10^{-1}$ | 1.7 | 1.9 |
| | 1.4 | $9.5 \times 10^{-2}$ | $1.3 \times 10^{-1}$ | $1.4 \times 10^{-1}$ | $1.9 \times 10^{-1}$ | $4.1 \times 10^{-1}$ | | 2.5 |
| | 1.6 | $1.3 \times 10^{-1}$ | $1.9 \times 10^{-1}$ | $2.2 \times 10^{-1}$ | $3.8 \times 10^{-1}$ | $6.2 \times 10^{-1}$ | | 3.9 |
| | 1.8 | $3.7 \times 10^{-1}$ | | | | | 6.3 | 7.4 |
| | 1.9 | | | | | | | |
| $\Delta s$ (pontos) | | 250 | 250 | 250 | 300 | 350 | 500 | 550 |
| $\Delta \xi$ (intervalo) | | $10^{-2}$ | $10^{-2}$ | $10^{-2}$ | $10^{-2}$ | $5 \times 10^{-3}$ | $10^{-3}$ | $10^{-3}$ |

Na figura 4, através da interpolação Spline Cúbica, fizemos os gráficos do tempo de processamento $T_p$, em função do parâmetro de relaxação $w$, para os vários valores considerados do parâmetro dielétrico $\beta$. Vê-se que para toda curva Spline os tempos de processamento ótimos são obtidos na proximidade de $w = 1.0$, independentemente dos valores de $\beta$, lembrando que a malha é refinada conforme $|\beta|$ cresce.

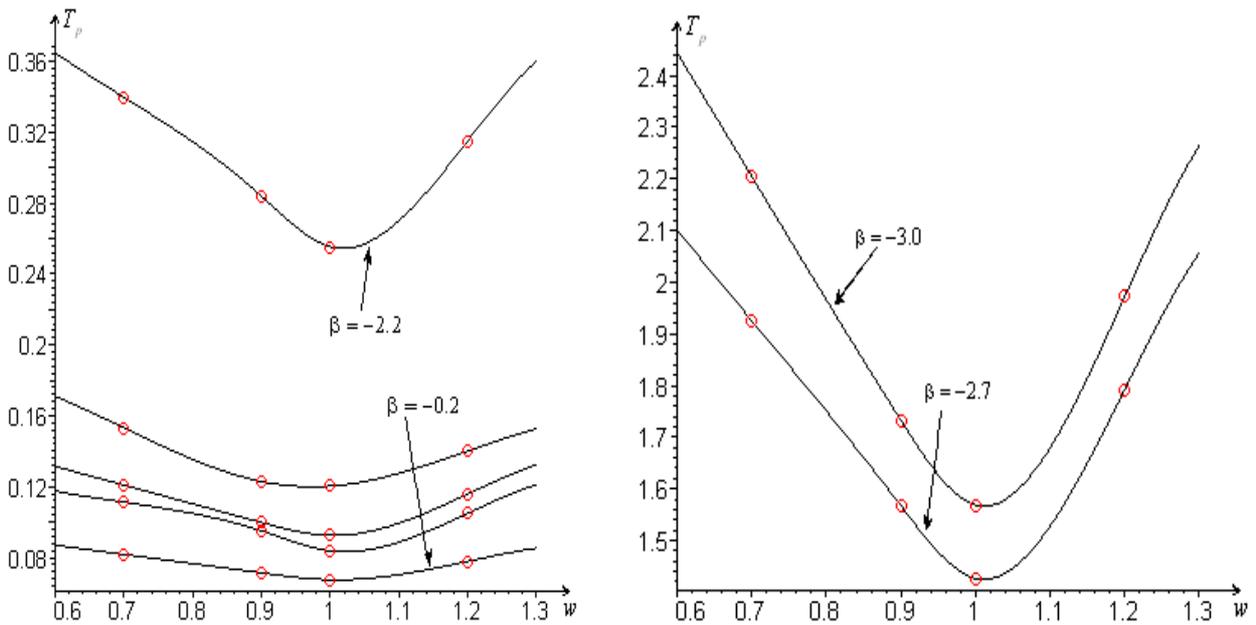

Figura 4: Interpolação Spline Cúbica para $T_p$ (minutos) como função de $w$, fixado $\beta$.

Outro resultado da tabela 1 pode ser visualizado na figura 5. Através do método dos mínimos quadrados, fazendo um ajuste de uma curva exponencial para os tempos de processamento quando $w = 1.0$, constatamos que o tempo de computação cresce com o aumento do $|\beta|$, consistentemente com De Oliveira et al. (2007).

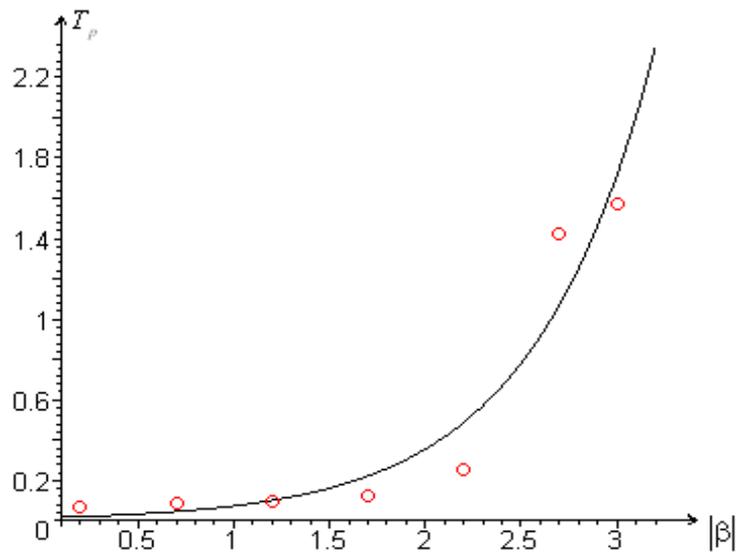

Figura 5: Ajuste exponencial para os valores mínimos de $T_p$ (minutos) em função de $|\beta|$.

## 5 Conclusões

Neste trabalho verificamos que o esquema numérico desenvolvido, fundamentado no método das diferenças finitas, mostrou-se relativamente simples do ponto de vista matemático-computacional, e adequado para a obtenção da solução sóliton em fibras óticas ideais.

Verificamos também que nosso esquema é sensível às variações do parâmetro $\beta$, como pode ser visualizado na figura 5, parâmetro que é uma medida da intensidade da não-linearidade do sistema de EDP's. Notamos que conforme o valor absoluto de $\beta$ cresce, para que a diferença entre as soluções analítica e numérica seja aceitável, devemos refinar a malha em $\Delta s$ e $\Delta \xi$, o que implicará inevitavelmente num maior tempo de processamento numérico.

Através de um processo de otimização numérica do tempo de simulação, obtivemos o parâmetro de relaxação ótimo com relação às variações do valor de $|\beta|$. A partir da figura 4, verifica-se que à medida que $w$ fica mais próximo de 1.0, as curvas Spline em $T_p \times w$, para os vários valores $\beta$ considerados, tendem ao seu tempo computacional mínimo.

Enfim, a partir destes resultados, podemos aplicar este esquema numérico otimizado para tratar o problema da propagação de sólitons em fibras óticas não-ideais, onde processos do tipo absorção, defeitos na fibra, ruídos, estão presentes, afetando a estabilidade das ondas sólitons.

## Agradecimentos

## Referências